\documentclass[%
 reprint,
nofootinbib,
 amsmath,amssymb,
 aps,
]{revtex4-1}

\usepackage{lipsum}
\usepackage{amsmath}
\usepackage{braket}
\usepackage{graphicx}
\usepackage{dcolumn}
\usepackage{bm}
\usepackage{hyperref}

\begin{document}

\title{Holographic heat engines and static black holes:\\ a general efficiency formula}

\author{Felipe Rosso}
\email{felipero@usc.edu}
\affiliation{%
Department of Physics and Astronomy, University of Southern California,\\ Los Angeles, California 90089-0484, USA
}%

\begin{abstract}
We study the efficiency of holographic heat engines in the context of extended black hole thermodynamics, where the cosmological constant becomes a dynamical variable. By taking the working substance as a static black hole (\textit{i.e.} a space-time with vanishing $C_V$) we derive an exact and analytic efficiency formula for virtually any engine defined by a cycle in the $p-V$ plane. This formula gives a simple criteria which completely resolves the benchmarking program for static black holes and shows that for any given engine there is an infinite family of tractable deformations which leave the efficiency invariant. We also derive an upper bound for the efficiency that holds for general engines.
\end{abstract}

\maketitle

\section{Introduction}

In recent times, the classic black hole thermodynamics \cite{Bardeen:1973gs,Bekenstein:1973ur,Hawking:1974sw} has been extended to include the cosmological constant $\Lambda$ as a dynamical pressure through the relation ${p=-\Lambda/8\pi}$. One of the main motivations for doing so is that in the presence of a non-vanishing cosmological constant the ordinary first law of black hole thermodynamics becomes inconsistent with the Smarr relation \cite{Smarr:1972kt}. This is resolved by taking $\Lambda$ as a dynamical variable, which implies that the mass of the black hole can no longer be identified with the internal energy but must be interpreted as the enthalpy \cite{Kastor:2009wy}
\begin{equation}\label{eq:1}
M=U+pV\ ,
  \qquad \qquad
  dM=TdS+Vdp\ ,
\end{equation}
where $V$ plays the role of the thermodynamic volume and is defined from the variaton of $M$ with respect to $p$. The dictionary for the temperature $T$ and entropy $S$ remains unchanged. This formalism provides with a natural extension of the ordinary thermodynamic and gives rise to several new and intereseting phenomena \cite{Cvetic:2010jb,Kubiznak:2012wp,Kastor:2014dra,Johnson:2014yja} (see Ref. \cite{Kubiznak:2016qmn} for a review). 

The notion of a holographic heat engine was first suggested in Ref. \cite{Johnson:2014yja}, by which a certain black hole is set in a thermodynamic cycle and produces work by changing its volume and pressure via ${dW=p\,dV}$. The engine is reffered as holographic given that it becomes particularly interesting in the context of the AdS/CFT correspondence \cite{Maldacena:1997re,Gubser:1998bc,Witten:1998qj} where the cycle in the $p-V$ plane corresponds to some type of flow between boundary CFTs \cite{Johnson:2014yja}. Though the precise mapping from the extended thermodynamic to the boundary CFT is not completely understood, there have been several interesting proposals \cite{Johnson:2014yja,Dolan:2014cja,Kastor:2014dra,Caceres:2015vsa,Caceres:2016xjz}. 

One of the central quantities characterizing a holographic heat engine is its efficiency $\eta=W/Q_h$, where $W$ and $Q_h$ are the total work and heat inserted into the system. From the second law of thermodynamics, this dimensionless quantity is upper bounded by Carnot's efficiency and provides with a classification of all possible cycles in the $p-V$ plane according to their proximity to this optimum value. It is therefore reasonable to expect the efficiency of a holographic heat engine to map into some important quantity characterizing the flow along the boundary CFTs.

The aim of this work is to provide an exact and analytic formula for the efficiency of an arbitrary holographic heat engine. We are able to do so by restricting to static black holes. The term static is used in the broader sense to refer to space-time geometries in which the horizon entropy $S$ and thermodynamic volume $V$ are not independent but determined from each other. We have in mind cases in which $S$ is determined from the horizon area and $V$ is given by the naive geometrical volume\footnote{A non-extensive set of examples of static black holes for which the results of this paper hold are: black holes in any space-time dimension of Einstein and Gauss-Bonnet gravity with spherical, planar and hyperbolic horizon charged under a Maxwell \cite{Johnson:2015ekr} or Born-Infeld \cite{Johnson:2015fva} sector. Though rotating black holes are in general non-static \cite{Cvetic:2010jb}, the rotating BTZ in three dimensions is static \cite{Hennigar:2017apu} (in the broader sense used in this paper).\label{ff}}.

For static black holes there are two known exact efficiency formulas, for a rectangular \cite{Johnson:2016pfa} and elliptical \cite{Chakraborty:2016ssb,Hennigar:2017apu} cycle in the $p-V$ plane. We generalize these results by deriving an exact efficiency formula that holds for virtually any holographic heat engine. The implications of this general formula are found to be far reaching. It gives a simple criteria that solves the benchmarking program of static black holes initiated in Ref. \cite{Chakraborty:2016ssb}, provides with a universal upper bound for the efficiency of a class of static black holes and has some remarkable consequences regarding the behavior of the efficiency under engine deformations.

This paper is organized as follows: we start in the following section by pointing out a simple yet powerful feature regarding heat flows along paths for static black holes. Using this observation we derive general bounds for the change in energy and adiabatic curves in the $p-T$ plane that will be useful when discussing efficiency bounds. The main results of the paper are presented in Sec. \ref{sec:3} where we derive the general efficiency formula and discuss its many implications. We finish in Sec. \ref{sec:5} with a summary of our results and a discussion of the difficulties of extending this procedure to non-static black holes.

\section{Constraint on heat flow of static black hole}
\label{sec:2}

\begin{figure}
\centering
\includegraphics[scale=0.5]{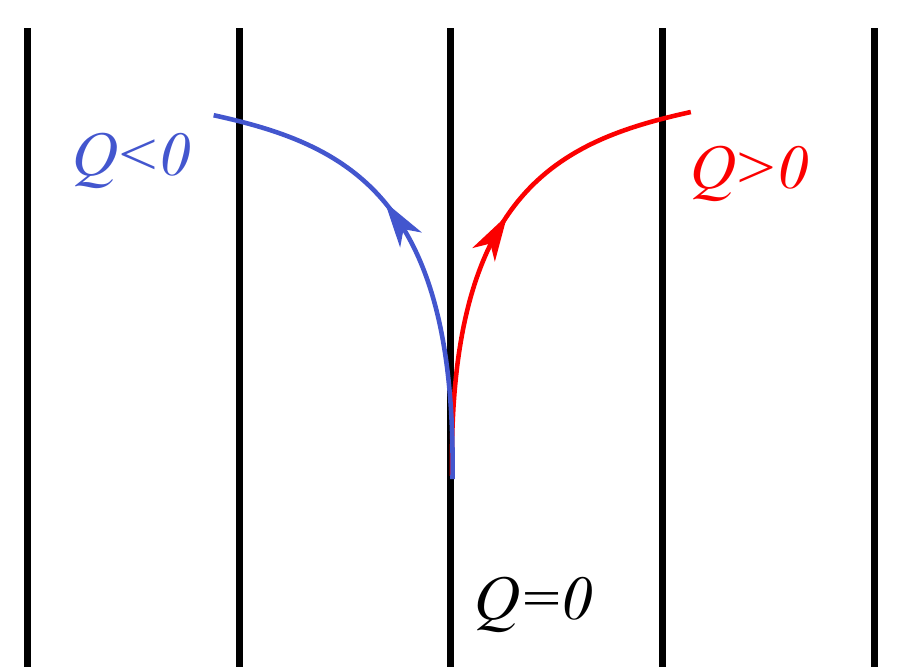}
\caption{Adiabatic curves in the $p-V$ plane for a static black hole, with two other paths which deviate from an adiabat to either side.}\label{fig:4}
\end{figure}

In this section we discuss a simple yet powerful feature of static black holes that will be central to our calculations. As discussed in the introduction, the term static black hole is used to refer to space-times in which the entropy $S$ and volume $V$ are determined from each other. This is the case for a wide range of black holes (see footnote \ref{ff}) in which $S$ is given by the horizon area and $V$ by the naive volume of the black hole. As first noted in Ref. \cite{Dolan:2010ha} this implies that an isochoric process $V={\rm const.}$ is equivalent to an adiabatic one $S={\rm const.}$ (in other words $C_V=0$). Therefore, the adiabatic curves for any static black hole are given by vertical lines in the \textit{whole} $p-V$ plane (see Fig. \ref{fig:4}).

Using this we can consider the following question: can we choose a reversible path in the $p-V$ plane so that the heat flow has a definite sign? To answer this consider the diagram of Fig. \ref{fig:4}, where we have plotted in black several vertical adiabats and two paths which deviate to either side. It is clear that the heat flow along the deviated paths is of opposite sign and different from zero, because they are not vertical. Since in the \textit{whole} $p-V$ plane the adiabats are given by vertical lines this means that any path that does not have a vertical section, by continuity cannot change the sign of its heat flow. The answer to our question is then the following: any path that can be described by a continuous \textit{function} (which by definition cannot have a vertical section) will have a fixed sign for its heat flow.

What about the precise sign of the heat flow in each direction? We can work this out by considering any specific path and checking its sign. Let us take an isothermic process going to the right in the $p-V$ plane. The heat exchanged is given by
$$Q_{T_0}=T_0\Delta S=T_0(A_{2}-A_1)/4\ ,$$ 
where $T_0$ is the temperature of the isotherm and we have assumed that the entropy is proportional to the area of the black hole.\footnote{Some static black holes may have other contributions to the entropy besides the area term, so that the argument presented here is not sufficient. If for a specific black hole it was found the opposite sign, most of our results follow with some appropriate changes.} If the volume of the black hole increases so does its area, which means that if the isotherm is covered from left to right, $V_2>V_1$ implies $A_2>A_1$ and therefore $Q_{T_0}>0$.

We can then conclude the following: any path in the $p-V$ plane described by a continuous function $f(V)$ and going from left to right, will have a positive heat flow into the system. If the path is covered in the opposite direction, it will have the inverse behavior. This simple property is central to this work and will have very interesting consequences.

Before moving on let us write an expression for the heat flow along a reversible path described by any function $f(V)$ connecting two arbitrary states. From the first law we can write the heat flow as $Q=\Delta U+W$, where $W$ is the work done by the black hole. Since the path is reversible we have $dW=f(V)dV$ so that $Q$ becomes
\begin{equation}\label{eq:14}
Q=\Delta U+\langle p \rangle \Delta V,
  \qquad \quad 
  \langle p \rangle=\frac{1}{\Delta V}\int_{V_1}^{V_2}dVf(V)\,,
\end{equation}
where we have defined the average pressure $\langle p \rangle$ along the path and $\Delta $ gives the difference between the final and initial state. Writing the work contribution in this manner will prove to be very natural and useful. Note that in general we can write the internal energy in terms of the mass of the black hole using Eq. (\ref{eq:1}).

\subsection{Bounds from heat flow constraint}
\label{sec:2.5}

We now explore the consequences of this simple property. Consider a path described by a function between any two points in the $p-V$ plane, so that the heat flow is given by Eq. (\ref{eq:14}). If we take the path from left to right the heat flow will be non-negative, which implies the following inequality
\begin{equation}\label{eq:15}
\frac{\Delta U}{\Delta V}\ge -\langle p \rangle \,,
\end{equation}
where we allow for the path to have vertical sections as long as it does not change its direction (these vertical sections do not contribute to the average pressure since $\langle p \rangle \propto W$).  For a path going in the opposite direction we get the same relation because both $\Delta U$ and $\Delta V$ get a minus sign and $\langle p \rangle$ is unchanged. This means that inequality (\ref{eq:15}) holds for any two points in the $p-V$ plane.

\begin{figure}
\centering
\includegraphics[scale=0.45]{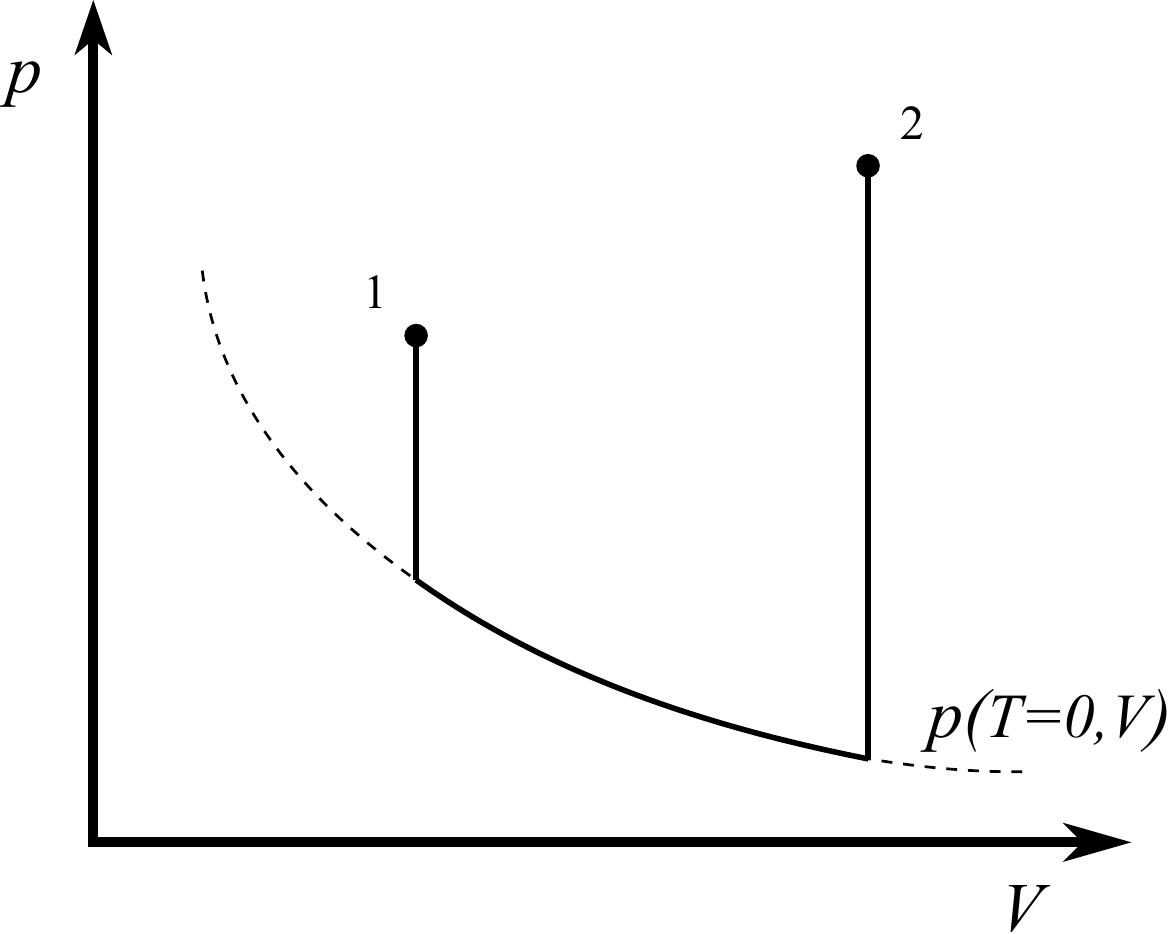}
\caption{Path between two arbitrary states which minimizes the average pressure. The dotted line corresponds to the pressure of the system at zero temperature $p(T=0,V)$. This will always be a non-increasing function of the volume for stable thermodynamic systems.}\label{fig:3}
\end{figure}

Notice that the right hand side is path dependent and the left side not. Since the inequality must hold for any path, we can choose a specific path which gives the most restrictive bound on the internal energy. Since the minimum pressure of any thermodynamic system is given at zero temperature, the critical path that minimizes the average pressure is the one shown in Fig. \ref{fig:3}.
The vertical sections do not contribute to the average pressure and the most restrictive version of Eq. (\ref{eq:15}) is given by
\begin{equation}\label{eq:12}
\frac{\Delta U}{\Delta V}\ge -\langle p \rangle_{T=0} =-\int_{V_1}^{V_2}dV\,\frac{p(0,V)}{\Delta V}\ .
\end{equation}
This inequality holds for any static black hole and will be useful for deriving an upper bound for the efficiency of holographic heat engines in the next section.

An interesting case follows when considering static black holes with vanishing pressure at zero temperature, so that $\langle p \rangle_{T=0}=0$ (this might also the case if the two points are at high enough volume, see Fig. \ref{fig:3}). For this class of black holes, Eq. (\ref{eq:12}) simplifies to
\begin{equation}\label{eq:18}
\frac{\Delta U}{\Delta V}\ge 0\,.
\end{equation}
We can find the black hole which saturates this bound by considering an AdS black hole in the limit of large volume, usually referred as an ``ideal gas'' black hole \cite{Johnson:2015ekr}, for which the mass is given by $M=pV$. Using this in Eq. (\ref{eq:1}) its internal energy vanishes $U=0$ and Eq. (\ref{eq:18}) becomes an equality. 

The bound (\ref{eq:18}) also becomes interesting when considering two states infinitesimally close to each other, so that we get a partial derivative instead. In particular, we can consider two points that have the same pressure and write the inequality in terms of the mass, so that Eq. (\ref{eq:18}) becomes
$$\left.\frac{\partial M}{\partial V}\right|_{p}\ge p\ .$$
Remarkably, this inequality can be integrated exactly using some of the usual thermodynamic tricks. Writing the differential of the mass (\ref{eq:1}) in terms of the variables $(p,V)$ we find
$$\left.\frac{\partial M}{\partial V}\right|_{p}=
  T\left.\frac{\partial S}{\partial V}\right|_{p}=
  T\left.\frac{\partial p}{\partial T }\right|_{S}
  \ge p \ ,$$
where in the second equality we have used one of Maxwell's thermodynamic identities. Integrating the expression we find the following bound
\begin{equation}\label{eq:17}
p\ge \left(p_0/T_0\right)T\ ,
\end{equation}
that holds for adiabatic paths and static black holes with $p(0,V)=0$ or in the large volume limit.

To understand this relation we consider an ``ideal gas'' black hole where it becomes an equality. In this case, the adiabatic vertical lines in the $p-V$ plane are directly mapped to linear curves in the $p-T$ plane with slope $p_0/T_0$ and passing through the origin. If we instead consider another black hole, inequality (\ref{eq:17}) provides a bound for the adiabatic curves in the $p-T$ plane. Though the adiabatic curves in the $p-V$ plane are extremely simple and given in Fig. \ref{fig:4}, their shape in the $p-T$ usually are not since the equation of state $p=p(T,V)$ might be quite complicated.

\section{General efficiency formula}
\label{sec:3}

The efficiency of an engine is defined as $\eta=W/Q_h$, where $Q_h$ is the heat flow into the system. Usually, one of the difficulties when calculating $\eta$ for an arbitrary substance and cycle, is that it is not easy to keep track of the signs of the heat flow along the paths to obtain $Q_h$. This is mainly because the adiabatic curves of the thermodynamic substance under consideration are non-trivial. However, when considering static black holes everything gets simplified since they have the simplest adiabatic curves of all: vertical lines in the whole $p-V$ plane. As discussed in Sec. \ref{sec:2} this implies that paths going to the right have positive heat flow and to the left negative, meaning that it is trivial to keep track of the sign of the heat flow along the cycle.

\begin{figure}
\centering
\includegraphics[scale=0.56]{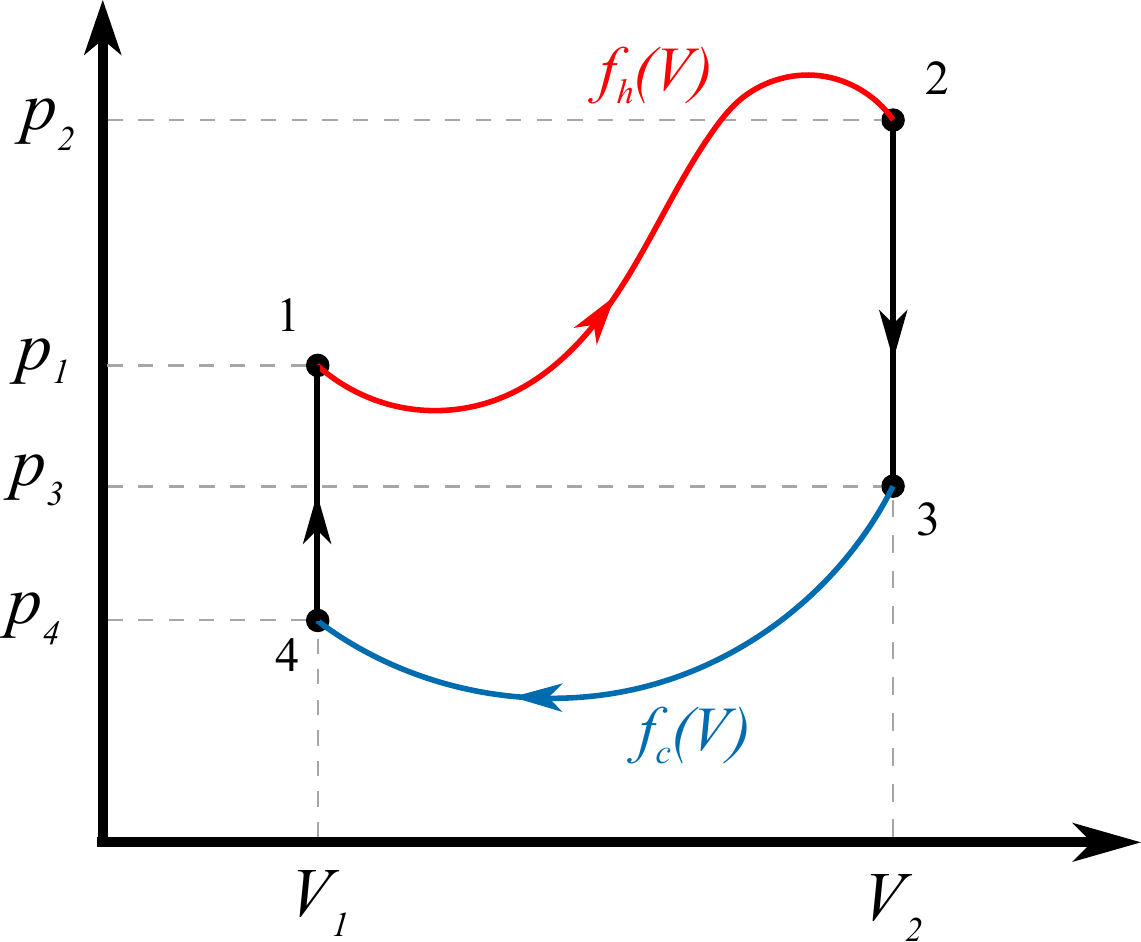}
\caption{General cycle defining a holographic heat engine. The process $2\rightarrow 3$ and $4 \rightarrow 1$ are given by isochors, equivalent to adiabats. The upper and lower paths are given by any pair of functions $f_h(V)$ and $f_c(V)$ respectively.}\label{fig:1}
\end{figure}

In the following, we will derive an exact efficiency formula for practically any heat engine defined by a cycle in the $p-V$ plane. To do this, we consider a heat engine defined by the general cycle in Fig. \ref{fig:1}.
The engine consists of two isochors (adiabats) and an upper and lower path given by \textit{any} pair of functions $f_h(V)$ and $f_c(V)$. We are actually considering a family of engines, determined by the freedom to choose these functions and the pressure and volume of the points $1-4$. Practically any reasonable heat engine can be obtained from this cycle.

Notice that there is a positive heat flow $Q_h$ into the engine \textit{only} along the upper path, and a negative flow $Q_c$ \textit{only} along the lower one. These heat flows can be written from Eq. (\ref{eq:14}) as
\begin{equation}\label{eq:3}
Q_h=
  (U_2-U_1)+\langle p \rangle_h(V_2-V_1)\ ,
\end{equation}
\begin{equation}\label{eq:5}
Q_c=(U_4-U_3)+\langle p \rangle_c (V_1-V_2)\ .
\end{equation}
where the subscript $h$ and $c$ corresponds to the average value along $f_h(V)$ and $f_c(V)$ respectively. To write the efficiency we use that $W$ is given by the area enclosed by the cycle, \textit{i.e.} ${W=(\langle p \rangle_h-\langle  p \rangle_c)(V_2-V_1)}$. Using this expression together with Eq. (\ref{eq:3}) in $\eta=W/Q_h$, we find the following efficiency formula 
\begin{equation}\label{eq:4}
\eta=\left(1-\frac{\langle p \rangle_c}{\langle p\rangle_h}\right)
     \left(\frac{\langle p \rangle_h}{\langle p \rangle_h+\frac{\Delta (M-pV)}{\Delta V}}\right)\,\ ,  
\end{equation}
where $\Delta$ is taken as the difference between the states $2$ and $1$, and we have written the internal energy explicitly in terms of the mass of the black hole using Eq. (\ref{eq:1}). This formula is our central result which holds for any static black hole and heat engine of the type given in Fig. \ref{fig:1}.

\subsection{Analysis and consequences of the efficiency formula}


To correctly understand the efficiency relation (\ref{eq:4}) it is important to keep in mind the difference between the specific working substance (black hole) under consideration and the characteristics of the heat engine, which is completely defined by the cycle in the $p-V$ plane. In this case, the heat engine from Fig. \ref{fig:1} is defined entirely by the pressure and volume of the points $1-4$, and the upper and lower paths. 

The first interesting feature of Eq. (\ref{eq:4}) is that its dependence on the particular black hole under consideration is extremely simple and only given by $\Delta M$. Every other factor depends on the definition of the engine. We can use this to solve the benchmarking program started in Ref. \cite{Chakraborty:2016ssb} which considers the following question: given a fixed engine is there any criteria that determines which black hole performs better? From Eq. (\ref{eq:4}) this question can be answered analitically and in complete generality: black holes with lower value of $\Delta M=M_2-M_1$ will result in a higher efficiency. It is quite remarkable that this general question has such a simple answer. 

In particular, this means that if there are two distinct black holes that have the same value of $\Delta M$, then their efficiency will be the same. This is the case when considering an asymptotically AdS charged black hole in Einstein and Gauss-Bonnet gravity. If the space-time dimension is equal to five, their masses only differ by a constant value \cite{Johnson:2015ekr}, which means that they will have the same $\Delta M$  and therefore the same efficiency on \textit{any} engine. Curiously this behavior is exclusive to the five dimensional case.

Now let us try to find an upper bound for the efficiency. From Eq. (\ref{eq:4}) we see that the efficiency is maximum, whenever $\Delta (M-pV)=\Delta U$ is minimum. Using inequality (\ref{eq:12}) derived in the previous section we find
\begin{equation}\label{eq:19}
\eta\le 
  \left(1-\frac{\langle p \rangle_c}{\langle p\rangle_h}\right)
     \left(\frac{\langle p \rangle_h}{\langle p \rangle_h-\langle p \rangle_{T=0}}\right)\,,
\end{equation}
where notice that both factors are non-negative. This is an upper bound which holds for any engine and static black hole. Different black holes will have different ${p(T=0,V)}$ curves and will result in distinct bounds. If we restrict to black holes with vanishing pressure at zero temperature the bound becomes
\begin{equation}\label{eq:16}
\eta\le 
  1-\frac{\langle p \rangle_c}{\langle p\rangle_h}=
  \eta_{\rm ideal\,\,gas}
\end{equation}
which is simply the efficiency of an ``ideal gas'' black hole (since $\Delta U=0$). This gives a simple and universal upper bound which holds for any heat engine described in Fig. \ref{fig:1} and black hole with $p(T=0,V)=0$.

\begin{figure}
\centering
\includegraphics[scale=0.45]{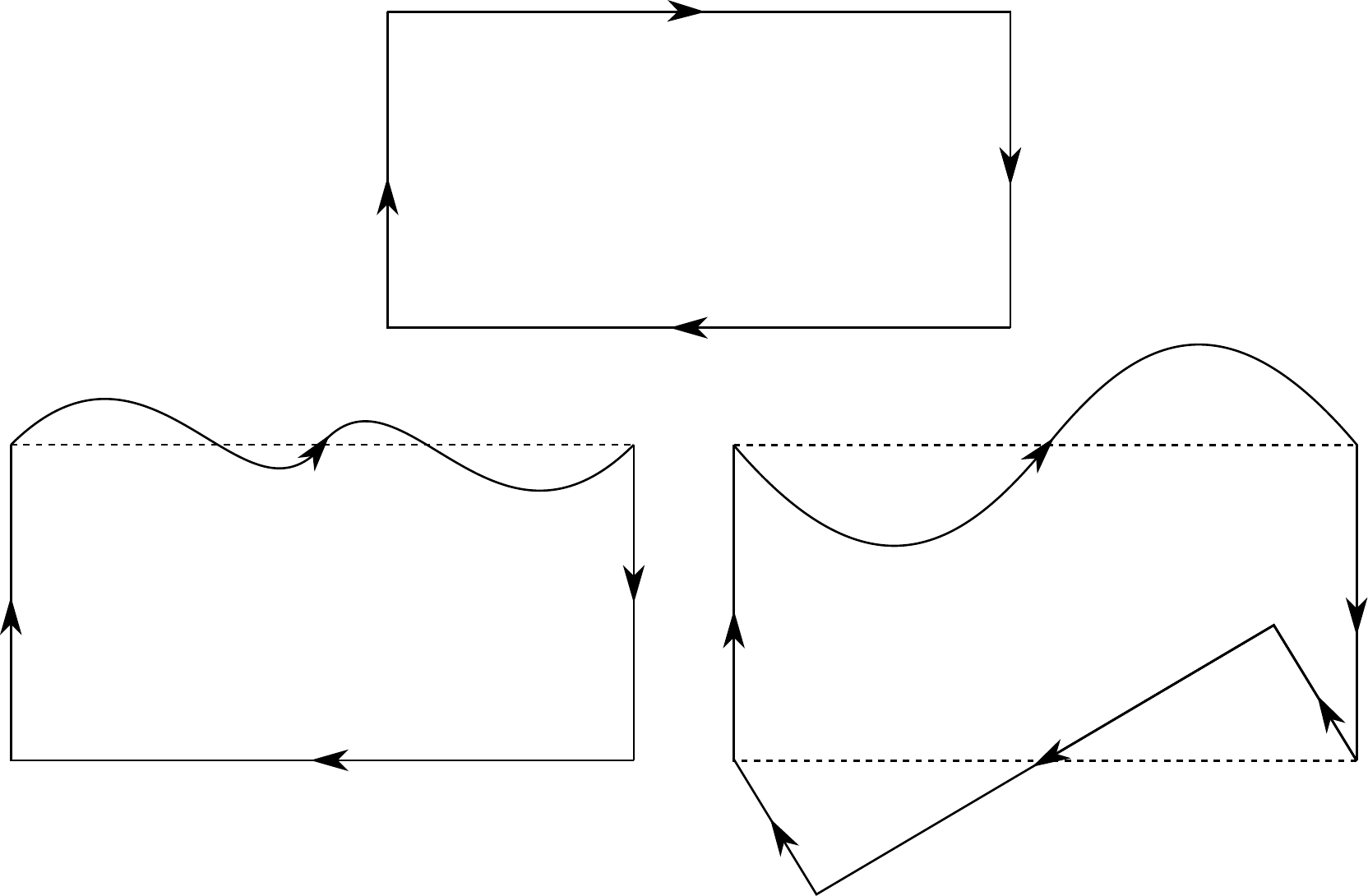}
\caption{Three very different engines that have the same efficiency, since the average pressure along the upper and lower paths is always given by $\langle p\rangle_h=p_1$ and $\langle p \rangle_c=p_4$.}\label{fig:5}
\end{figure}

Finally, let us analyze the dependence of Eq. (\ref{eq:4}) under deformations of the engine. Notice that apart from the coordinates of the four points $1-4$, the formula only depends on the averages of the upper and lower paths $f_h(V)$ and $f_c(V)$. Apart from being rather simple, this means that two very different engines will have the same efficiency as long as their averages along those paths is the same. In Fig. \ref{fig:5} we sketch some examples of engines which have the same efficiency despite of being radically different. This means that every efficiency calculation performed in the literature for a given cycle in the ${p-V}$ holds not only for that particular engine but for an infinite family of deformations analogous to the ones in Fig. \ref{fig:5}. This uncovers a remarkable and previously unknown behavior regarding holographic heat engines and static black holes.

\subsection{Application to particular engines}
\label{sec:4}

We can now apply the efficiency formula in Eq. (\ref{eq:4}) to particular heat engines. Let us start by recovering some previously know examples and then move to consider new engines.

\subsubsection{Rectangular engine}

We can get a rectangular engine from the cycle in Fig. \ref{fig:1} by taking $p_1=p_2$, $p_3=p_4$ and the upper and lower paths as isobars, which means that the average pressures are given by $\langle p \rangle_h=p_1$ and $\langle p \rangle_c=p_4$. Using this in Eqs. (\ref{eq:3}) and (\ref{eq:5}) and writing everything in terms of the black hole mass we find
\begin{equation}\label{eq:20}
\eta_{\rm rectangular}=1-\frac{|Q_c|}{Q_h}=1-\left(\frac{M_3-M_4}{M_2-M_1}\right)\,,
\end{equation}
which agrees with the result obtained in Ref. \cite{Johnson:2016pfa}. If we consider an ``ideal gas'' black hole in Eq. (\ref{eq:16}) we get ${\eta=1-p_4/p_1}$, in agreement with the calculations in Refs. \cite{Johnson:2014yja,Johnson:2015ekr}. 

Most efficiency calculations in the literature have been carried out for this rectangular cycle. The new insight is that the formula in Eq. (\ref{eq:20}) and all the previous calculations do not only hold for the rectangular engine, but for an infinite number of deformations as the ones sketched in Fig. \ref{fig:5}. 

\subsubsection{Elliptical engine}

To get an elliptical engine from the general cycle in Fig. \ref{fig:1}, we consider $p_i=p$ for $i=1,\dots,4,$ and the upper and lower paths as half ellipses centered at ${((V_1+V_2)/2,p)}$. Taking the horizontal radius of the ellipse as ${R_v=(V_2-V_1)/2}$ and the vertical one as $R_p$, the average pressures can be easily computed as
$$\langle p \rangle_h=p+\frac{\pi}{4} R_p\ ,
  \qquad \qquad
  \langle p \rangle_c=p-\frac{\pi}{4}R_p\,.$$
Using this in Eq. (\ref{eq:4}) we get the efficiency of an elliptical engine
\begin{equation}\label{eq:7}
\eta_{\rm elliptical}=
    \frac{2}
  {1+2\left(\frac{\Delta M}{\pi R_p R_v}\right)}\,,
\end{equation}
which agrees with the result in Ref. \cite{Hennigar:2017apu} and in Ref. \cite{Chakraborty:2016ssb} for the ideal gas case. If we compare this formula with the calculations of Ref. \cite{Chakraborty:2016ssb}, where the efficiency was computed numerically for three different static black holes, we find agreement to the third significant figure. Same as before, Eq. (\ref{eq:7}) holds for any deformation of the cycle which leaves the average pressure along the paths unchanged.

\subsubsection{Axially symmetric engine}

\begin{figure}
\centering
\includegraphics[scale=0.52]{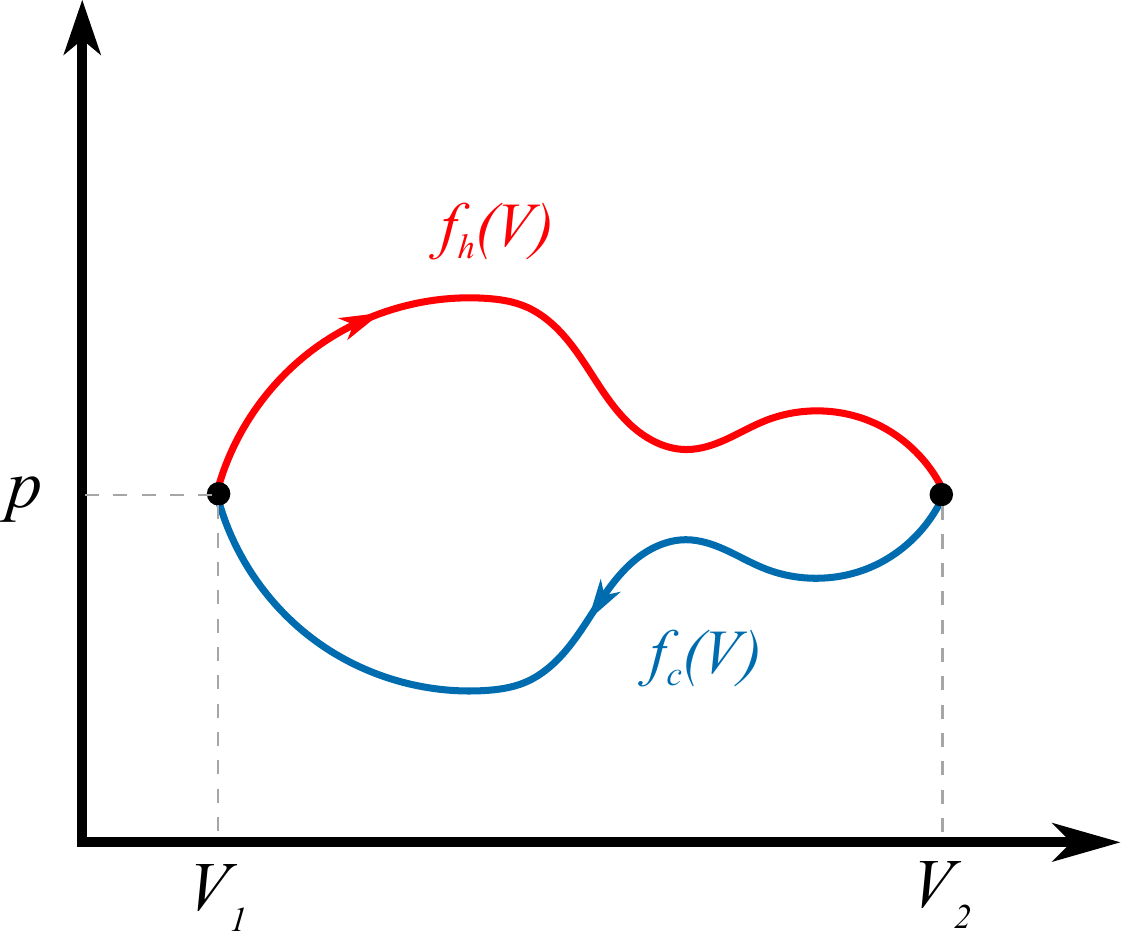}
\caption{Axially symmetric heat engine. The upper path can be given by any function $f_h(V)$, while the lower one is fixed by $f_c(V)=2p-f_h(V)$, so that the cycle is axially symmetric.}\label{fig:2}
\end{figure}

There is an interesting feature of the efficiency formula for an elliptical engine in Eq. (\ref{eq:7}): it only depends on the ratio of $\Delta M$ and the area enclosed by the engine, which is equal to the work $W$. Are there other types of engines which display this simple behavior? To answer this, consider a cycle that is symmetric with respect to a horizontal axis passing through its center, as we see in Fig. \ref{fig:2}.

Comparing with the general cycle in Fig. \ref{fig:1} we can get this engine by taking $p_i=p$ for $i=1,\dots,4,$ and fixing the lower path according to $f_c(V)=2p-f_h(V)$, so that the cycle is axially symmetric. The function $f_h(V)$ remains unconstrained. A quick calculation shows that the average pressures are given by
\begin{equation}\label{eq:8}
\langle p \rangle_c=2p-\langle p \rangle_h
  \qquad \qquad
  \langle p \rangle_h=
  p+\frac{W}{2\Delta V}\, ,
\end{equation}
so that from the general efficiency formula in Eq.  (\ref{eq:4}) we find
\begin{equation}\label{eq:10}
\eta_{\rm symmetric}=
  \frac{2}
  {1+2\left(\frac{\Delta M}{W}\right)}\,,
\end{equation}
which provides a nice generalization of Eq. (\ref{eq:7}). 

\subsubsection{Triangular engine}

Another interesting and simple engine is given by considering the axially symmetric engine from Fig. \ref{fig:2}, but taking either the upper or lower path as an isobar. When the isobar is taken along the lower path we will regard the triangular engine as being positive $(+)$, while name it negative $(-)$ when it is the other way. 

For a positive triangular engine the average pressures are given by $\langle p \rangle_c=p$ and
$\langle p \rangle_h=p+W/\Delta V$. Comparing with the expression in Eq. (\ref{eq:8}) for $\langle p \rangle_h$ there is a difference in a factor of two because in this case the cycle is half the size. Using these relations in Eq. (\ref{eq:4}) we get the efficiency for a positive triangular engine
$$
\eta_{\rm triangular}^{(+)}=\frac{1}
  {1+\left(\frac{\Delta M}{W}\right)}\,.
$$
  
For a negative triangular engine the average pressures are given by $\langle p \rangle_c=p-W/\Delta V$ and $\langle p \rangle_h=p$, so that the efficiency is equal to
$$
\eta_{\rm triangular}^{(-)}=
  \frac{W}
  {\Delta M}\,.
$$
This expression is particularly simple due to the fact that the upper path, which is the only part which contributes to $Q_h$ is an isobar.

We could continue calculating efficiencies for many other engines, but at this point we decide to stop, since the main features and techniques have already been exposed.

\section{Final remarks}
\label{sec:5}

In this work, we have exploited the fact that static black holes have very simple adiabatic curves to derive an exact and analytic efficiency formula for any holographic heat engine defined by a cycle in the $p-V$ plane. Using this formula we have solved the benchmarking program for static black holes, whose goal is to rank the black holes according to how well they perform under a fix engine. The simple criteria we have found is that for the general engine in Fig. \ref{fig:1} the black hole with larger efficiency will be the one with lower value of ${\Delta M=M_2-M_1}$. The question remains open for non-static black holes where one must usually resort to numerical techniques \cite{Chakraborty:2017weq,Hennigar:2017apu}. Additionally, we obtained the upper bound in Eq. (\ref{eq:19}) which becomes universal for black holes with vanishing pressure at zero temperature (\ref{eq:16}). For this class of black holes the bound is saturated by the ``ideal gas'' black hole. 

Maybe the most remarkable consequence of our calculations is the conclusion that any deformation of the general engine in Fig. \ref{fig:1} which leaves the average pressures unchanged does not modify the efficiency. Apart from giving an equivalence relation between an infinite number of cycles in the $p-V$ plane, this means that all the efficiency formulas derived in Sec. \ref{sec:4} are much more general than previously thought. For instance, the simple expression in Eq. (\ref{eq:20}) for a rectangular engine is also valid for the complicated cycles in Fig. \ref{fig:5}. This means that all the calculations performed in the literature for the efficiency of static black holes on rectangular engines are valid for much more general cycles. 

A natural question that arises is whether these methods can be extended to include non-static black holes. The difficulty facing when wanting to make progress in this direction is that there is no general formula for the adiabatic curves of non-static black holes. This means that the simple observation that paths going to the right have a positive heat inflow is not valid anymore. Then, if we try to calculate the efficiency of an engine like the one from Fig. \ref{fig:1}, the upper and lower paths may have sections with positive and negative contributions of heat, so that we are unable to keep track of the direction of the heat flow and get expressions for $Q_h$ or $Q_c$. Nevertheless, it would be interesting to try to calculate the adiabatic curves for a simple non-static black hole and apply a similar construction.

\section*{Acknowledgments}

Felipe Rosso thanks Clifford V. Johnson, Avik Chakraborty and Scott MacDonald for comments on the manuscript.

\bibliography{Efficiency}

\end{document}